\documentclass[12pt]{utarticle}
\usepackage{cite}
\usepackage{hyperref}
\numberwithin{equation}{section}
\input xy
\xyoption{all}

\def\clap#1{\hbox to 0pt{\hss#1\hss}}

\begin{document}
\thispagestyle{empty}

{\parskip 0ex
\hfill VT-IPNAS 10-12

\hfill TCC-022-10                                        

\hfill UTTG-08-10%
}

\vspace{1in}

\begin{center}

{\large\bf Quantization of Fayet-Iliopoulos Parameters}

{\large\bf in Supergravity}

\vspace{0.1in}

Jacques Distler$^1$, Eric Sharpe$^2$

\begin{tabular}{cc}
{ \begin{tabular}{c}
$^1$ University of Texas, Austin\\
Department of Physics\\
Austin, TX  78712-0264
\end{tabular} }
&
{ \begin{tabular}{c}
$^2$ Physics Department\\
Virginia Tech \\
Blacksburg, VA  24061
\end{tabular} }
\end{tabular}

\email{\tt distler@golem.ph.utexas.edu},
\email{\tt ersharpe@vt.edu}\\

$\,$

\end{center}

In this short note we discuss quantization of the Fayet-Iliopoulos parameter
in supergravity theories.  We argue that in supergravity, the
Fayet-Iliopoulos parameter determines a lift of the group action to
a line bundle, and such lifts are quantized.
Just as D-terms in rigid $\mathcal{N}=1$ supersymmetry
are interpreted in terms of moment maps and symplectic reductions,
we argue that in supergravity the quantization of the Fayet-Iliopoulos
parameter has a natural understanding in terms of linearizations in
geometric invariant
theory (GIT) quotients, the algebro-geometric version of symplectic
quotients.

\begin{flushleft}
July 2010
\end{flushleft}

\newpage
\thispagestyle{empty}
\tableofcontents
\vfill
\newpage
\setcounter{page}{1}

\section{Introduction}

The recent paper \cite{nati0} discussed quantization of Fayet-Iliopoulos
parameters in four-dimensional 
supergravity theories in which the group action on
scalars was realized linearly.  In this short note, we observe that 
that quantization condition has a more general understanding,
as a choice of lift of the group action to a line bundle over
the moduli space.  Such a lift is precisely a
inearization in the sense of geometric invariant theory
(GIT) quotients, the algebro-geometric analogue of symplectic quotients.

After reviewing the result that K\"ahler classes on moduli spaces
in supergravity are integral forms in section~\ref{bag-ed-rev}, 
we discuss the quantization of the Fayet-Iliopoulos parameter in 
section~\ref{fi-quant}.  In rigid supersymmetry, the Fayet-Iliopoulos
parameter is interpreted in terms of symplectic quotients, and is
not quantized.  We argue that in ${\cal N}=1$ supergravity in
four dimensions, the Fayet-Iliopoulos
parameter should instead be interpreted in terms of a choice of
lift of the $G$ action to a holomorphic line bundle,
and such choices are quantized.  In section~\ref{git-interp},
we discuss how to interpret the supergravity quotient in terms of the
algebraic-geometry version of symplectic quotients, known as 
geometric invariant theory quotients.
In section~\ref{susy-break} we briefly comment on implications of
this work for discussions of supersymmetry breaking in supergravity.
Finally in section~\ref{higher-susy} we conclude with some observations
on analogues in ${\cal N}=2$ supergravity in four dimensions.
In appendices we discuss pertinent sigma model anomalies, conditions
for bundles to admit lifts of group actions, and
work through a simple example of a geometric invariant theory quotient.

The recent paper \cite{nati0} also discussed two-dimensional
theories defined by restricting sums over instantons to a subset of
all instantons.  Such theories are the same as strings on gerbes,
special kinds of stacks, as is discussed in the physics literature
in for example
\cite{kps,nr,msx,glsm,karp1,karp2,hhpsa,cdhps}
and reviewed in conference proceedings including
\cite{me-vienna,me-tex,me-qts}.
(There is also a significant mathematics literature on Gromov-Witten
invariants of stacks and gerbes; see
for example \cite{cr,agv,cclt,mann} for a few representative examples
of that literature.)
A more direct description of a string on a gerbe is as the (RG endpoint of)
a gauged sigma model in which the group acts ineffectively, meaning a
subgroup acts trivially.  More globally, gauging ineffective group actions
and restricting nonperturbative sectors go hand-in-hand.

In the special case of stacks that are gerbes, {\it i.e.} the theories
discussed in \cite{nati0}, such theories in two dimensions are
equivalent to nonlinear sigma models on disjoint unions of
spaces \cite{hhpsa}, a result named the ``decomposition conjecture.''
We can understand the decomposition conjecture schematically as follows.
Consider a nonlinear sigma model on a space $X$, for simplicity with
$H^2(X,{\bf Z}) = {\bf Z}$, with a restriction on
worldsheet instantons to degrees divisible by $k$.  We can realize that
restriction in the path integral by inserting a projection operator
\begin{displaymath}
\frac{1}{k} \sum_{n=0}^{k-1} \exp\left( i \int \phi^* \left( 
\frac{2 \pi n}{k} \omega \right) \right)
\end{displaymath}
where $\omega$ is the de Rham image of a generator of $H^2(X,{\bf Z})$.
Inserting this operator into a partition function is equivalent to working
with a sum of partition functions with rotating $B$ fields,
and this is the essence of the decomposition conjecture.

One of the applications of the result above is to Gromov-Witten theory,
where it has been checked and applied to simplify computations of
Gromov-Witten invariants of gerbes, see \cite{ajt1,ajt2,ajt3,t1,gt1,xt1}.
Another application is to gauged linear sigma models \cite{cdhps}, where it
answers old questions about the meaning of the Landau-Ginzburg point
in a GLSM for a complete intersection of quadrics, gives a physical
realization of Kuznetsov's homological projective duality \cite{kuz1,kuz2,kuz3},
and updates
old lore on GLSM's.

A more detailed discussion of gerbes, including four-dimensional
theories (which have somewhat different properties from the two-dimensional
ones reviewed above), examples of gerby moduli `spaces' in field and
string theory, and a discussion of the Fayet-Iliopoulos quantization
condition for gerby moduli spaces in supergravity,
will appear in~\cite{hs}.

\section{Review of Bagger-Witten}   \label{bag-ed-rev}

Bagger and Witten \cite{bag-ed1}
discussed how the K\"ahler class of the moduli space\footnote{
The arguments of \cite{bag-ed1}, and our own arguments here, all assume
that the moduli space of the supergravity theory is a smooth manifold.
} 
of scalars
of a supergravity theory is quantized, resulting from the fact that
ultimately the K\"ahler class must be the first Chern class of a
line bundle over the moduli space.
In this section we will describe an analogous argument for
quantization of the Fayet-Iliopoulos term in supergravity.

Let us briefly begin by reviewing the quantization of
Newton's constant in ungauged supergravity theories,
following \cite{bag-ed1}.
First, across coordinate patches on the moduli space,
the K\"ahler potential $K$ transforms as
\begin{displaymath}
K \: \mapsto \: K \: + \: f \: + \: \overline{f}
\end{displaymath}
where $f$ is a holomorphic function of moduli.
To be a symmetry of the theory, this must be accompanied by a
rotation of the gravitino $\psi_{\mu}$ 
and the superpartners $\chi^i$ of the
scalar fields on the moduli space:
\begin{equation}  \label{grav-trans}
\chi^i \: \mapsto \: \exp\left( + \frac{i}{2} {\rm Im}\, f
\right) \chi^i, \: \: \:
\psi_{\mu}  \: \mapsto \: \exp\left( - \frac{i}{2}{\rm Im}\, f
\right) \psi_{\mu}
\end{equation}
(Since the $\chi^i$ and $\psi_{\mu}$ are chiral fermions, these are
chiral rotations, hence there are potential anomalies -- see
for example \cite{dist-blog} or appendix~\ref{app-anom}
for a discussion.)

Consistency of the rotations~(\ref{grav-trans}) across triple overlaps
(even within classical physics) implies that
the $f$'s define a line bundle with even $c_1$, to which the
fermions $\chi^i$, $\psi_{\mu}$ couple.
In more formal language, if we let that line bundle be
$\mathcal{L}^{\otimes 2}$, we can summarize~(\ref{grav-trans}) by
saying that the gravitino is a spinor-valued section of $TX \otimes
\phi^* \mathcal{L}^{-1}$, where $X$ is the four-dimensional low-energy effective
spacetime and $\phi: X \rightarrow  M$ the boson of the four-dimensional
nonlinear sigma model on the compactification moduli space 
$ M$, and that the fermions $\chi^i$ are spinor-valued sections of
$\phi^*(T  M \otimes \mathcal{L})$.
In the same language, the K\"ahler form on $ M$
is a (de Rham representative of)
$c_1(\mathcal{L}^{- 2})$ (and hence an even integral form).
Moreover, given how the $\chi^i$ transform, 
the superpotential $W$ transforms as a holomorphic
section of $\mathcal{L}^{\otimes 2}$.
Because the K\"ahler form determines the metric on the
fermi kinetic terms, which must be positive-definite,
\cite{bag-ed1} argues that $\mathcal{L}^{\otimes 2}$ must be a 
negative bundle -- so if the moduli space $M$ is a smooth compact
manifold, then the superpotential must vanish.

\section{Quantization of the Fayet-Iliopoulos parameter}
\label{fi-quant}

Now, let us imagine gauging a group action on the target space
$ M$ of the nonlinear sigma model above.
We will argue that in supergravity, one must lift the
group action on the base $M$ to the line bundle ${\cal L}$,
and that the Fayet-Iliopoulos
parameter corresponds to such a choice of lift.  As there are
integrally-many choices of lifts, possible values of the
Fayet-Iliopoulos parameter are quantized.

Let us begin by reviewing how one gauges group actions in nonlinear
sigma models in general.
To preserve supersymmetry (see {\it e.g.}
\cite{bag-ed3,bag-w}), the group action must be generated
by holomorphic Killing vectors
\begin{displaymath}
X^{(a)} \: \equiv \: X^{(a)i} \frac{\partial}{\partial \phi^i}
\end{displaymath}
where $(a)$ denotes a Lie algebra index, and $\phi$ a map in the
nonlinear sigma model.
To be holomorphic Killing means they must satisfy
\begin{equation*}
\begin{split}
\nabla_i X^{(a)}_j \: + \: \nabla_j X^{(a)}_i & = 0 \\
\nabla_{ \overline{\imath} } X^{(a)}_j \: + \:
\nabla_j X^{(a)}_{\overline{\imath}} & = 0
\end{split}
\end{equation*}
On a K\"ahler manifold, the first equation holds automatically.
The second equation implies that there exist real scalar functions
$D^{(a)}(\phi^i, \phi^{\overline{\imath}})$ such that
\begin{equation}\label{MomentMaps}
\begin{split}
g_{i \overline{\jmath}} X^{(a) \overline{\jmath}} & = \: i 
\frac{\partial}{\partial \phi^i} D^{(a)} \\
g_{i \overline{\jmath}} X^{(a) i} & = \: - i \frac{
\partial}{\partial \phi^{\overline{\jmath}} } D^{(a)}
\end{split}
\end{equation}
These conditions only determine the $D^{(a)}$ up to additive constants.

Quantities $D^{(a)}$ solving the equations above 
are known as ``Killing potentials,'' and are moment maps for the
group action \cite{bag-ed3,bag-w}.  
Because only their derivatives are defined,
they are ambiguous up to a constant shift,
and such constant shifts are Fayet-Iliopoulos parameters.

In rigid supersymmetry, we interpret the gauging mathematically
as a symplectic quotient in symplectic geometry.
The $D^{(a)}$ define moment maps, and the constant shifts,
the Fayet-Iliopoulos parameters, define the coadjoint orbit on
which the symplectic reduction takes place.
For a gauged $U(1)$, say, there is a single Fayet-Iliopoulos parameter
which can take any real value, defining symplectic quotients with
symplectic forms in real-valued cohomology.

In supergravity, however, that picture is problematic,
as can be seen from the following quick and slighty sloppy argument.
The value of the Fayet-Iliopoulos parameter determines the K\"ahler form
on the quotient, but as we just outlined,
\cite{bag-ed1} have argued that in an ungauged
moduli space, 
the K\"ahler form is integral
(and even).
To get an integral K\"ahler form on the quotient, the
Fayet-Iliopoulos parameter must 
be quantized.

The simplest example is the construction of $\mathbb{P}^n$ as
a symplectic quotient of $\mathbb{C}^{n+1}$ by $U(1)$.  One begins on
$\mathbb{C}^{n+1}$ with an integral (in fact trivial)
K\"ahler form, but by varying the image
of the moment map, one can recover $\mathbb{P}^n$ with any
real K\"ahler class, not necessarily integral.
To get an integral K\"ahler class, the image of the moment map must
also be integral.  More generally \cite{allenpriv}, 
for abelian $G$, the moment map takes
values in $t^*$, but only if one reduces on points in $T^* \subset t^*$
can one hope to get an integral K\"ahler form on the quotient.

Hence, to get an integral K\"ahler form on the quotient,
the Fayet-Iliopoulos parameter must be quantized.

This argument is a little too slick; 
it is not immediately obvious, from the arguments of \cite{bag-ed1}, 
that the K\"ahler class on the symplectic quotient must also be quantized. 
In the case of linearly-realized group actions, as was recently 
discussed in \cite{nati0}[section 3], it is easy to make the 
argument precise.  The supergravity actions contains
\begin{displaymath}
-3 \int d^4 \Theta \, E \exp(-K - r V)/3
\end{displaymath}
where $r$ is the Fayet-Iliopoulos parameter.
As a result, gauge transformations
\begin{displaymath}
V \: \mapsto \: V \: + \: \Lambda \: + \: \overline{\Lambda}
\end{displaymath}
act as K\"ahler transformations with $f = - r \Lambda$.
Thus, the gauge symmetry acts as an R-symmetry under which the superpotential
has charge $-r$.  As a result, if a superfield $\Phi^j$ has charge $q_j$,
then the fermion $\chi^j$ has charge $q_j + r/2$, and so charge
quantization implies that $r/2$ must be an integer,
{\it i.e.} the Fayet-Iliopoulos parameter must be an even integer.

However, it remains to understand this problem more generally. 
What is clear from the geometrical discussion of \S\ref{bag-ed-rev} is that, 
for the purposes of the supergravity theory, it does not suffice to 
define the action of $G$ on $M$; we also need an equivariant lift\footnote{
Given $\{ g \in G \}$, such a lift is a set $\{ \tilde{g} \}$ acting
on the bundle such that $\tilde{g} \tilde{h} = \widetilde{gh}$.
These are known technically as a $G$-equivariant structure or linearization.  
See
appendix~\ref{app-equiv-exist} for technical details and remarks on their
existence.
} of the 
$G$-action to the line bundle $\mathcal{L}$, and as we shall discuss
later, such equivariant lifts, when they exist,
are quantized.  In particular, we will identify
Fayet-Iliopoulos parameters with, in essence, a choice of equivariant
lift, and this is the ultimate reason for their quantization in
supergravity.

We can see Fayet-Iliopoulos parameters as lifts
explicitly in the supergravity 
lagrangians of \cite{wb}[chapter 25].
In general, since the fermions $\chi^i$, $\psi_{\mu}$ couple
to $\mathcal{L}$, $\mathcal{L}^{-1}$, a group action on $M$ must be 
lifted
to an
action on $\mathcal{L}$, $\mathcal{L}^{-1}$ in order to 
uniquely define the theory.
(A group action on either of $\mathcal{L}$, $\mathcal{L}^{-1}$ defines a 
group action on the other, so henceforth we will only speak about
group actions on $\mathcal{L}$.)
We can see infinitesimal lifts explicitly in the infinitesimal group actions
for real $\epsilon^{(a)}$ \cite{wb}[(25.14)]:
\begin{equation*}
\begin{split}
\delta \phi^i & = \: \epsilon^{(a)} X^{(a) i} \\
\delta A_{\mu}^{(a)} & =  \partial_{\mu} \epsilon^{(a)} \: + \:
f^{abc} \epsilon^{(b)} A_{\mu}^{(c)} \\
\delta \chi^i & = \: \epsilon^{(a)} \left(
\frac{\partial X^{(a) i} }{\partial \phi^j} \chi^j \: + \:
\frac{i}{2} {\rm Im} \: F^{(a)} \chi^i \right) \\
\delta \lambda^{(a)} & =\: f^{abc} \epsilon^{(b)} \lambda^{(c)}
\: - \: \frac{i}{2} \epsilon^{(a)} {\rm Im} \: F^{(a)} \lambda^{(a)} \\
\delta \psi_{\mu} & = \: - \frac{i}{2} \epsilon^{(a)} {\rm Im} \: F^{(a)}
\psi_{\mu}
\end{split}
\end{equation*}
where $F^{(a)} = X^{(a)} K + i D^{(a)}$ ($K$ the K\"ahler potential), 
and $F^{(a)}$ is easily checked to be
holomorphic.
For real $\epsilon^{(a)}$,
the K\"ahler potential undergoes a standard K\"ahler transformation
\begin{displaymath}
\delta K \: = \: \epsilon^{(a)} F^{(a)} \: + \: 
\epsilon^{(a)} \overline{F}^{(a)}
\end{displaymath}
hence in the gauge transformations above,
terms proportional to ${\rm Im}\: F^{(a)}$ are precisely encoding
the K\"ahler transformations on fermions given in 
equation~(\ref{grav-trans}).
Thus, the gauge-transformation terms proportional to ${\rm Im} \: F^{(a)}$ 
(also known as super-Weyl transformations) 
appear to encode an infinitesimal lift 
of the group action to $\mathcal{L}$.  Strictly speaking, infinitesimal
lifts are required to obey the Lie algebra:
\begin{equation}  \label{inf-lift}
\left[ \delta^{(a)}, \delta^{(b)} \right] \psi_{\mu} \: = \:
 \frac{i}{2} \epsilon^{(a)} \epsilon^{(b)} f^{abc} {\rm Im}\: F^{(c)}
\psi_{\mu}
\end{equation}
(for real $\epsilon^{(a)}$).
The $D^{(a)}$ can be chosen to obey \cite{wb}[equ'n (24.6)]
\begin{displaymath}
\left[ X^{(a) i} \partial_i \: + \:
X^{(a) \overline{\imath} } \partial_{\overline{\imath}} \right]
D^{(b)} \: = \:
- f^{abc} D^{(c)}
\end{displaymath}
and it is straightforward to check that with this choice,
the $F^{(a)}$ do indeed satisfy equation~(\ref{inf-lift}), and hence
define an infinitesimal lift of $G$ (equivalently, an equivariant lift
of the Lie algebra).
Shifts in the imaginary part of $F^{(a)}$
are precisely Fayet-Iliopoulos parameters, hence, Fayet-Iliopoulos parameters
encode a choice of equivariant lift.

Next, we shall show that the allowed values of the Fayet-Iliopoulos
parameter are constrained (in fact, quantized) by the condition
that the infinitesimal lifts integrate to honest (global) lifts.
As noted above, the infinitesimal group action on ${\cal L}$,
the infinitesimal lift, is given by
\begin{displaymath}
+ \frac{i}{2} \epsilon^{(a)} {\rm Im}\, F^{(a)}
\end{displaymath}
so that the lift of the group element
\begin{displaymath}
g \: \equiv \: \exp\left( i \epsilon^{(a)} T^{a} \right)
\end{displaymath}
($T^a$ generators of the Lie algebra) is
\begin{displaymath}
\tilde{g} \: \equiv \: \exp\left( \frac{i}{2} \epsilon^{(a)} {\rm Im}\, F^{(a)} \right)
\end{displaymath}
We require that the group be represented honestly, not projectively,
{\it i.e.} $\tilde{g} \tilde{h} = \widetilde{gh}$,
in order to define an honest lift of the group $G$
(known technically as a $G$-equivariant structure or in this case,
a $G$-linearization).
Shifting the Fayet-Iliopoulos parameters (translating
$D^{(a)}$) corresponds to a 
$g$-dependent rescaling of $\tilde{g}$:
\begin{equation*}
\tilde{g} \: \mapsto \: \tilde{g} \exp\left( i \theta_g \right)
\end{equation*}
since $F^{(a)} = X^{(a)} K + i D^{(a)}$.
We can use such shifts to try to produce an honest representation
if the $\tilde{g}$'s do not already form an honest representation;
nevertheless, we might not be able to do so.
Let the group formed by the $\tilde{g}$ be denoted $\tilde{G}$,
then for real\footnote{
For algebraic groups, we have the nearly identical sequence
\begin{equation*}
1 \: \longrightarrow \: \mathbb{C}^{\times} \: \longrightarrow \:
\tilde{G}_{ \mathbb{C} } \: \longrightarrow \:
G_{\mathbb{C}} \: \longrightarrow \: 1.
\end{equation*}
} Lie groups we have a short exact sequence
\begin{equation}\label{gexactseq}
1 \: \longrightarrow \: U(1) \: \longrightarrow \:
\tilde{G} \: \longrightarrow \: G \: \longrightarrow \: 1
\end{equation}
We can shift Fayet-Iliopoulos parameters to get an honest representation
if and only if the extension $\tilde{G}$ splits as $G \times U(1)$.
In general, this will not always be the case -- equivariant
structures lifting group actions do not always exist. As we explain in 
appendix~\ref{app-equiv-exist}, an equivariant moment map \eqref{inf-lift} suffices to guarantee an equivariant lift of the Lie algebra. For $G$ connected and simply-connected, we show that this suffices to guarantee that \eqref{gexactseq} splits and gives an euivariant lift of $G$. 

Assuming that we can find an honest representation, 
{\it i.e.} assuming an honest lift exists,
we still have a little freedom left in the Fayet-Iliopoulos parameters:
we can deform $\tilde{g}$'s by $\theta_g$'s that represent $G$.
In other words, if 
\begin{displaymath}
\theta_g \: + \: \theta_h \: = \: \theta_{gh}
\end{displaymath}
for all $g, h \in G$, then we can shift the Fayet-Iliopoulos parameters
to give phases as 
\begin{displaymath}
\tilde{g} \: \mapsto \: \tilde{g} \exp\left(i \theta_g \right)
\end{displaymath}
while maintaining an honest representation:
\begin{eqnarray*}
\left( \tilde{g} \exp\left( i \theta_g \right) \right)
\left( \tilde{h} \exp\left( i \theta_h \right) \right)
& = &
\tilde{g} \tilde{h} \exp\left( i \left( \theta_g \: + \: \theta_h \right)
\right) \\
& = & \widetilde{gh} \exp\left( i \theta_{gh} \right)
\end{eqnarray*}
Such shifts $\theta$ (i.e., the \emph{difference} between two splittings of \eqref{gexactseq}) are classified by ${\rm Hom}(G, U(1))$
(for real Lie groups $G$) or ${\rm Hom}( G_{\bf C}, {\bf C}^{\times} )$
(for algebraic groups $G_{\bf C}$).  These are, clearly, the only remaining
allowed Fayet-Iliopoulos shifts.

For example, if the gauge group is $U(1)$, then
we can shift
\begin{displaymath}
\frac{1}{2} {\rm Im}\, F^{(a)} \: \pm \: \left( {\rm integer} \right)
\end{displaymath}
and leave the group representation invariant.
This quantized shift is the Fayet-Iliopoulos parameter.

The fact that honest lifts, when they exist, are quantized in the
fashion above,
is a standard result in the mathematics literature
(see {\it e.g.} \cite{kostant1}[prop. 1.13.1]).
Since it also forms the intellectual basis for
the central point of this paper, let us give
a second
explicit argument  
that differences
between lifts are quantized, following\footnote{
Essentially the same argument, in a different context, is responsible for
understanding discrete torsion as a choice of equivariant structure 
on the $B$ field.
} \cite{dt3}.
Assume the space is connected,
and let $\{ U_{\alpha} \}$ be an open cover, that is `compatible' with the
group action\footnote{We omit details concerning covers.  The result
whose derivation we are sloppily outlining here is standard.
}.
At the level of Cech cohomology, a $G$-equivariant line bundle
is defined by transition functions $g_{\alpha \beta}$,
a gauge field $A_{\alpha}$, and related
data such that
\begin{equation*}
\begin{split}
g^* A_{\alpha} & = \: A_{\alpha} \: + \: d \ln h^g_{\alpha} \\
g^* g_{\alpha \beta} & = \: (h^g_{\alpha}) (g_{\alpha \beta})
(h^g_{\beta})^{-1} \\
h^{g_1 g_2}_{\alpha} & = \: (g_2^* h^{g_1}_{\alpha})(h^{g_2}_{\alpha}) 
\end{split}
\end{equation*}

Now, suppose we have two distinct equivariant structures, two lifts,
defined by $h^g_{\alpha}$ and $\overline{h}^g_{\alpha}$.
Define
\begin{displaymath}
\phi^g_{\alpha} \: \equiv \: \frac{ h^g_{\alpha} }{ \overline{h}^g_{\alpha} }
\end{displaymath}
From the consistency condition on $g^* g_{\alpha \beta}$ for each
equivariant, we find that $\phi^g_{\alpha} = \phi^g_{\beta}$,
{\it i.e.} $\phi^g_{\alpha}$ is the restriction to $U_{\alpha}$ of
a function we shall call $\phi^g$.
From the consistency condition on $g^* A_{\alpha}$ for each
equivariant structure, we find that $\phi^g$ is a locally constant function,
and finally from the remaining consistency condition we find that
\begin{displaymath}
\phi^{g_1 g_2} \: = \: \phi^{g_2} \phi^{g_1}
\end{displaymath}
{\it i.e.} $\phi$ defines a homomorphism $G \rightarrow U(1)$.
In other words, on a connected manifold,
the difference\footnote{
Note that the lifts themselves cannot be canonically identified
with elements of ${\rm Hom}(G, U(1))$ -- unless
the line bundle is trivial, there is in general no canonical ``zero''
lift --- {\it i.e.}, 
there is no natural "zero" for the Fayet-Iliopoulos parameter.
Instead, the {\it set} of linearizations is acted upon
freely by this group. Technically, we say the set of linearizations is a torsor under this group.
} between any two $G$-lifts 
is an
element of ${\rm Hom}(G,U(1))$.

If the gauge group is $U(1)$, then as
${\rm Hom}(U(1), U(1)) = \mathbb{Z}$,
we see that the difference between
any two lifts is an integer.

Applying to the present case, we find that the difference between
any two versions of
\begin{displaymath}
\frac{1}{2} {\rm Im} \: F^{(a)}
\end{displaymath}
is an integer, and since ${\rm Im} \: F = D + \cdots$,
we see that the difference between any two allowed values of
the Fayet-Iliopoulos parameter must be an even
integer.  (Any {\it e.g.} one-loop counterterm would merely
product an overall shift; the difference between any two
allowed values would still be an integer.)

\section{Interpretation in geometric invariant theory}
\label{git-interp}

In rigid supersymmetry, the $D$ terms and Fayet-Iliopoulos parameters
are interpreted in terms of symplectic quotients and symplectic reduction.
In supergravity, however, there are some key differences:
\begin{enumerate}
\item As we saw in the previous section, in supergravity the
Fayet-Iliopoulos parameter is quantized, because it acts as a lift
of the group action on $M$ to a line bundle, $\mathcal{L}$.  
These structures have no analogue
in rigid supersymmetry.
\item A more obscure but also important point is that 
$\mathcal{L}$ defines a projective embedding of $M$.
The quantization of the K\"ahler form $\omega$ described in \cite{bag-ed1}
means that $M$ is a Hodge manifold.
By the Kodaira Embedding Theorem \cite{kodaira1}, every Hodge manifold is
projective, and $\mathcal{L}^{-n}$, for some $n\gg 0$ is the ample line bundle
that provides the projective embedding.
\end{enumerate}
These features are not characteristic of symplectic quotients,
but they are characteristic of the algebro-geometric analogue of
symplectic quotients, known as geometric invariant theory (GIT)
quotients (see {\it e.g.}
\cite{git,newstead,kirwan}) instead.

In a GIT quotient, the analogue of the image of the
moment map is quantized.  Briefly,
when both are defined, GIT quotients are essentially
equivalent to symplectic quotients, except that GIT quotients are constructed
to always have integral K\"ahler classes, whereas symplectic quotients
can have arbitrary real K\"ahler classes.

In a GIT quotient, instead of quotienting the inverse image of the
moment map by a real Lie group, one considers instead a quotient
of a complex manifold by an algebraic group (typically the complexification
of a real Lie group, which is very natural from the point of view of 
$\mathcal{N}=1$ supersymmetry). 
The effect of quotienting by a complex algebraic group
turns out to be functionally equivalent to first taking the inverse
image of the moment map then quotienting by a real Lie group
(a subgroup of the algebraic group).
In GIT quotients, the quotient is built via an explicit embedding
into a projective space (technically, the quotient is constructed as
Proj of a graded ring of group invariants), and the K\"ahler class on the
quotient is the pullback along that embedding of the first
Chern class of ${\cal O}(1)$.
To build a GIT quotient, we must specify
a lift (technically, a `linearization')
of the (algebraic) group action to a line bundle $L$ on the
space being quotiented, whose first Chern class is the K\"ahler class
upstairs.
As this technology may not be widely familiar to physicists,
in appendix~\ref{git-ex} we show explicitly how projective spaces can be
constructed in this form.

We have seen these structures in $\mathcal{N}=1$ supergravity --
the moduli space $M$ has a projective embedding by virtue
of $\mathcal{L}^{-1}$, and in order to define the quotient we
must pick a $G$-linearization of $\mathcal{L}^{-1}$.
In fact, the Fayet-Iliopoulos parameter is understood in
terms of such a choice of lift of the group action.
Thus, the structure of gaugings in $\mathcal{N}=1$ supergravity 
closely parallels the key features of GIT quotient constructions.

That said, some of the technical details of GIT quotient
constructions are rather different.  In a GIT quotient, for
example, quotients are build via
embeddings into projective spaces constructed from invariant rings
(thus the name), whereas symplectic reductions are built as
$G$-quotients of fibers of the moment map.
In the present case, although we see projective embeddings and
Fayet-Iliopoulos parameters as $G$-linearizations of an ample
line bundle,
the rest of the quotient construction more nearly follows the
standard symplectic story ($G$-quotients of fibers of a moment map)
rather than that of geometric invariant theory (as
invariant coordinate rings are not completely central, modulo the
discussion above).

The construction of the GIT quotient, in terms of the ring of invariant functions, is closely reminiscent of the approach to four-dimensional gauge theories, where one describes the moduli space in terms of its (invariant) chiral rings (see for
example \cite{argyres-lect}[section 12.3] and references therein, though also see \cite{Aspinwall:1994uj} for a different perspective).

Perhaps the best interpretation of the D-terms in supergravity is that
the Fayet-Iliopoulos parameter is defined by a choice of linearization,
though the rest of the construction should still be interpreted
in terms of symplectic quotients.  In particular, a choice of linearization
directly defines a moment map.  We can see this as follows.
Let $G$ act on a space $X$, which is lifted to a linearization on a
line bundle $L$ over $X$.  Suppose $G$ preserves a connection one-form $A$
on
$L$, whose curvature is the symplectic form $\omega$.
Then pairing vector fields from Lie($G$) with
$A$ gives real-valued functions on $X$, in the usual fashion:
\begin{displaymath}
i_{V_g} \omega \: = \: d \mu_g
\end{displaymath}
for a function $\mu: X \times {\rm Lie}(G) \rightarrow \mathbb{R}$.
Such a pairing is equivalent to a moment map $X \rightarrow {\rm Lie}(G)^*$,

\section{Supersymmetry breaking}
\label{susy-break}

A sufficient condition for supersymmetry breaking in supergravity is
that $\langle D^{(a)} \rangle \neq 0$.  One result of our analysis is
that, in principle, for some moduli spaces and line bundles ${\cal L}$,
there may not exist an allowed translation of $D^{(a)}$,
for which $\langle D^{(a)} \rangle = 0$.  In such a case, supersymmetry
breaking would be inevitable.

An example of this phenomenon is discussed in \cite{Bagger:1982ab}[section 5].
There, $M = \mathbb{P}^1$. The group of isometries is $SO(3)$, 
but when $\mathcal{L}$ is an odd power of the tautological line bundle, 
the group that has an equivariant lift is actually $G = SU(2)$. 
Since ${\rm Hom}(SU(2),U(1))$ is trivial, the equivariant lift is unique.

Moreover for $\mathcal{L}= \mathcal{O}(-n)$, 
\begin{equation*}
  (D^{(1)})^2 + (D^{(2)})^2 + (D^{(3)})^2 = \left(\frac{n}{2\pi}\right)^2
\end{equation*}
independent of the location on $\mathbb{P}^1$.  Supersymmetry is always broken.

As another example, consider gauging just a $U(1)$ subgroup of the isometry group of $M = \mathbb{P}^1$. The allowed moment maps are
\begin{equation*}
D = -\frac{1}{2\pi} \left(\frac{n}{1+|\phi|^2} + k\right)
\end{equation*}
for any $k\in \mathbb{Z}$. 
Different choices of $k$ correspond to different allowed values of the 
Fayet-Iliopoulos coefficient. There are two fixed points of the $U(1)$ action, the north pole ($\phi=0$) and the south pole ($\phi'=1/\phi=0$). 
Both are extrema of the scalar potential (minima, for an appropriate range of 
$k$). 
For generic choice of $k$, supersymmetry is broken at both points. 
For $k=-n$, supersymmetry is unbroken at the north pole, 
and broken at the south pole. 
For $k=0$, supersymmetry is broken at the north pole, 
and unbroken at the south pole. 
Exchanging the roles of north and south pole requires shifting $k\to -n-k$, 
reflecting the fact that there's no canonical ``zero" for the FI coefficient; 
rather, they form a  torsor for ${\rm Hom}(G,U(1))$. 

\section{Higher supersymmetry}   \label{higher-susy}

We have not investigated higher supersymmetry cases thoroughly,
though we will make some basic observations regarding
${\cal N}=2$ supergravity in four dimensions.
For example, consider the hypermultiplet moduli space.
In rigid ${\cal N}=2$ supersymmetry, that moduli space is a hyperK\"ahler
manifold, but in ${\cal N}=2$ supergravity it is a quaternionic
K\"ahler manifold \cite{bag-ed2}.
It was argued in \cite{bagger-ssm}[equ'n (5.16)]
that in ${\cal N}=2$ supergravity in four dimensions, the
curvature scalar on the quaternionic K\"ahler moduli manifold
is uniquely determined, so that there is not even an integral
ambiguity.  Similarly, it seems to be a standard result that
in quaternionic K\"ahler reduction, unlike hyperK\"ahler reduction,
there is no Fayet-Iliopoulos ambiguity in the moment map, but rather
the moment map is uniquely defined\footnote{
Our intuition for this is that in ${\cal N}=2$ supergravity, there is
a triplet of Fayet-Iliopoulos parameters, which (from the discussion of
this section) must all be integral, and yet can also be rotated under
the action of an $SU(2)_R$.  The only triple of integers consistent with
$SU(2)_R$ rotation is $(0,0,0)$.
} \cite{galicki1,swann1,joyce1}.

One can consider also the moduli space of vector multiplets in
${\cal N}=2$ supergravity.
Such moduli spaces are described by special geometry, and in this
case
(see {\it e.g.} \cite{strom}[equ'n (10)]) 
the K\"ahler form on the moduli space arising in
Calabi-Yau compactifications is identified with 
\begin{displaymath}
\partial \overline{\partial} \ln \langle \Omega | \overline{\Omega} \rangle
\end{displaymath}
and hence unique (as this is invariant under rescalings of the holomorphic
top-form $\Omega$).
All of these results tell us that
triplets
of $D$ terms in ${\cal N}=2$ supergravity
have no Fayet-Iliopoulos ambiguity. 

We leave a careful
analysis of ${\cal N}=2$ supergravity to future work.

\section{Conclusions}

In this paper we have reviewed recent discussions of quantization of
the Fayet-Iliopoulos parameter in supergravity theories.
We argued that,
In general, that quantization can be understood formally via a choice
of linearization on a line bundle appearing in the theory, linking
gauging in supergravity models with `geometric invariant theory'
quotients.

The recent paper \cite{nati0} went one step further to consider
{\it e.g.} $U(1)$ gaugings with nonminimal charges, which
(as argued in the introduction) correspond mathematically to
sigma models on gerbes.  The claims of \cite{nati0} regarding such
models can be understood as arising from the fact that there are more
(`fractional') line bundles over gerbes than exist over the underlying
spaces.  We will consider such models in the upcoming work 
\cite{hs}.

\section{Acknowledgements}

We would like to thank D.~Freed,
S.~Hellerman, A.~Knutson, T.~Pantev, and M.~Verbitsky for
useful conversations.  E.S. would also like to thank the
University of Pennsylvania for hospitality while this work
was completed.

J.D. was partially supported by the US-Israel Binational
Science Foundation and NSF grant PHY-0455649.
E.S. was partially supported by NSF grants
DMS-0705381 and PHY-0755614.

\appendix

\section{Four-dimensional sigma model anomalies}  \label{app-anom}

In section~\ref{bag-ed-rev} we described the classical physics of the
fermions in an ${\cal N}=1$ supergravity theory in four dimensions.
Because those fermions are chiral, and hence (necessarily) undergo
chiral rotations across coordinate patches, there is the potential
for an anomaly.

In this appendix we will briefly outline how the resulting anomalies,
following \cite{dist-blog} (see also \cite{mn1,mn2,mmn} 
for background information
on sigma model anomalies).

Globally, the fact that the fermions undergo chiral rotations across
coordinate patches is encoded in an anomaly, given by the six-form
piece of
$\hat{A}(X) \wedge {\rm ch}(E)$,
where
\begin{displaymath}
E \: = \: \phi^* \left(T  M \otimes \mathcal{L} \right)
\: \ominus \:
\left( TX \ominus 1 \right) \otimes \phi^* \mathcal{L}^{-1}
\end{displaymath}
(The first term is from the superpartners $\chi^i$ of the chiral superfields,
the second from the gravitino $\psi_{\mu}$.)
In the notation above, $M$ is the target space of the nonlinear
sigma model (the moduli space of the supergravity theory),
$X$ is the four-dimensional spacetime, $\phi: X \rightarrow  M$
the boson of the nonlinear sigma model, and $\mathcal{L}$ is the line bundle
encoding the chiral rotations across overlaps.

The six-form piece above (for the $E$ appropriate for supergravity) is given by
\begin{multline}\label{anom}
\phi^* {\rm ch}_3( M) \: - \:
\frac{1}{24} p_1(X) \phi^* c_1( M)  \: + \:
\phi^* c_1(\mathcal{L}) \left( \phi^* {\rm ch}_2( M)\: + \:
\frac{21-n}{24} p_1(X) \right)\\
\: + \:
\frac{1}{2} \phi^* \left( c_1(\mathcal{L})^2 c_1( M) \right) \: + \:
\frac{n+3}{6} \phi^* c_1(\mathcal{L})^3
\end{multline}
where $n$ is the number of chiral superfields, the dimension of the
moduli space $M$.

The first two terms are independent of $\mathcal{L}$, and are present even 
in the case of rigid supersymmetry. 
They rule out many classically-sensible supersymmetric sigma models, 
for instance projective spaces of dimension greater than two. 
Of course, for the phenomenologically-interesting case of the sigma model 
which arises for a spontaneously-broken global symmetry, 
$G\to H$, $M= T^*(G/H)$, the anomaly, in the rigid case, vanishes.

When coupled to supergravity, with nontrivial $\mathcal{L}$, the anomaly 
polynomial takes the above, more complicated, form. 
The moduli spaces which arise in string theory are typically noncompact but 
can, nonetheless, have quite complicated topology. 
So \eqref{anom} seems to provide a nontrivial constrain. 
Unfortunately, these moduli spaces are typically not smooth varieties, 
but rather are stacks. 
The extension of these considerations, to the case of $M$ a stack, 
will be pursued elsewhere.

For present purposes, we would like to extend \eqref{anom} to the case where 
we gauge some global symmetry, $G$, of $M$. 
Two things change, when we do this. 
First of all, $\phi$ is no longer a map from $X$ to a fixed manifold, $M$. 
Rather, let $P\to X$ be a $G$-principal bundle. 
We form the associated bundle
\begin{equation*}
    \mathcal{M} = (P\times M)/G
\end{equation*}
with fiber $M$. Now, $\phi$ is a \emph{section}
\begin{displaymath}
\xymatrix{
{\mathcal M} \ar[d] \\
X \ar@/^1pc/[u]^{\phi}
}
\end{displaymath}

The first effect of this change is to replace $TM$ in the above expression, 
with $T_{\text{vert}}\mathcal{M}$. 
The second change is that, in supergravity, the gaugini transform as 
sections of $\phi^*\mathcal{L}^{-1}$. 
The net effect is to modify \eqref{anom} to
\begin{multline}\label{anom2}
\phi^* {\rm ch}_3(T_{\text{vert}}\mathcal{M}) \: - \:
\frac{1}{24} p_1(X) \phi^* c_1(T_{\text{vert}}\mathcal{M}) \\
 \: + \:
\phi^* c_1(\mathcal{L}) \left( \phi^* {\rm ch}_2(T_{\text{vert}}\mathcal{M}) +
\frac{21\: -\: n\: + \: {\rm dim}(G)}{24} p_1(X) \right)\\
\: + \:
\frac{1}{2} \phi^* \left( c_1(\mathcal{L})^2 c_1(T_{\text{vert}}\mathcal{M})
 \right) \: + \:
\frac{n \: + \: 3 \: - \: {\rm dim}(G)}{6} \phi^* c_1(\mathcal{L})^3
\end{multline}

Even when \eqref{anom2} does not vanish, it is possible to contemplate 
cancelling the anomaly by adding to the action "Wess-Zumino"-type terms, 
whose classical variation is anomalous \cite{Wess,WittenGlobal}, 
but we will not pursue that here.

\section{Existence of equivariant structures}
\label{app-equiv-exist}

As noted in the text, $G$-equivariant structures on line bundles do not
always exist.  In this appendix, we shall work out conditions for finding
such equivariant structures.

In general, if a group $G$ acts on a manifold $M$, then a $G$-equivariant
structure on a line bundle $L$ is a lift of $G$ to the total space
of that line bundle, {\it i.e.} for all $g \in G$, a map $\tilde{g}:
{\rm Tot}(L) \rightarrow {\rm Tot}(L)$ such that
\begin{displaymath}
\pi(gy) \: = \: g \pi(y)
\end{displaymath}
for all $y \in {\rm Tot}(L)$,
and such that $\tilde{g} \tilde{h} = \widetilde{gh}$.
Furthermore, one often imposes additional constraints,
{\it e.g.} an equivariant lift that preserves a holomorphic structure
or a connection.

It is a standard result in the mathematics literature
that equivariant structures do not always exist.
The obstruction is typically finding a lift such that
$\tilde{g} \tilde{h} = \widetilde{gh}$ -- often one can find a projective
representation of the group, but it may not be possible to find an honest
representation of the group.

One necessary condition for equivariant structures to exist is that
characteristic classes be invariant under group actions, but this is
not sufficient. Examples of non-equivariant line and vector bundles, with invariant
characteristic classes, can be found in {\it e.g.}
\cite{Sharpe:2009hr,Donagi:2003tb} and references therein.
Another example is as follows\footnote{
We would like to thank T.~Pantev for providing this example.
}.
Let $E$ be an elliptic curve with a marked point $\sigma \in E$.
Let $L = \mathcal{O}_{E}(2\sigma)$.  Let $x \in E$, and let
$t_x: E \rightarrow E$ be the translation by $x$ in the group law.
Then for any $x$, the automorphism $t_x$ preserves $c_1(L)$ simply because
$t_x$ is homotopic to the identity and so acts trivially on
$H^2(E, \mathbb{Z})$.  However, if $x$ is a general point, $t_x^* L$
is not isomorphic to $L$ as a holomorphic bundle.  Hence this group of 
translations on $E$ clearly can not have an equivariant lift\footnote{
In fact, $t_x^* L 
\cong L$ if and only if $x$ is a point of order two on $E$.
Now, even if we restrict to points of order two, we will not have an
equivariant structure.  The points of order two are an abelian group
isomorphic to $\mathbb{Z}_2 \times \mathbb{Z}_2$ which preserves $L$, but $L$
does not have an equivariant structure.
}.
However, this sort of group action is not of interest, even in 
globally-supersymmetric sigma models (let alone locally-supersymmetric). 
There is no moment map for the translation action. 
(In the language of \eqref{theseq}, below, the vector field generating this 
symmetry is not in the kernel of $s$.)

In \S\ref{fi-quant}, we noted the condition for an equivariant lift 
(or linearization\footnote{
Technically,
a `linearization' is an equivariant structure in which the group acts linearly
on the fibers of vector bundles:  $L_x \rightarrow L_{gx}$.
All of the equivariant structures appearing in this paper are examples
of linearizations.
}, in the nomenclature of GIT
quotient constructions) of the infinitesimal $G$-action. 
Here, we will lay out those conditions more carefully, and consider 
the existence of an equivariant lift for finite $G$ transformations.

The K\"ahler form, $\omega$, endows $M$ with a symplectic structure.
(To be precise, we will use $\omega' = \tfrac{1}{2} \omega$ as the
symplectic structure.) 
At the Lie algebra level, $\mathfrak{g}\subset \mathcal{X}_H$,
the Lie algebra of Hamiltonian vector fields on $M$; in other words,
the $\mathfrak{g}$-action arises from globally-defined 
moment maps \eqref{MomentMaps}. 
There is an exact sequence, 
\begin{equation}\label{theseq}
   0\: \longrightarrow \: H^0(M,\mathbb{R}) \: \longrightarrow \: 
 C^\infty(M) \: \longrightarrow \: \mathcal{X}_\omega 
\overset{s}{\longrightarrow} H^1(M,\mathbb{R})\: \longrightarrow \: 0
\end{equation}
where $\mathcal{X}_{\omega'}$ is the Lie algebra of symplectic vector fields
(those preserving $\omega'$), and 
$\mathcal{X}_H\subset \mathcal{X}_{\omega'}$ is the subalgebra
which is the kernel of $s$.
So we naturally have a central extension
\begin{equation}\label{galgextension}
0\: \longrightarrow \:
 H^0(M,\mathbb{R})\: \longrightarrow \:
 \tilde{\mathfrak{g}}\: \longrightarrow \: \mathfrak{g} \: \longrightarrow \: 0
\end{equation}
where $\tilde{\mathfrak{g}}$ acts equivariantly on $\mathcal{L}$
(an infinitesimal lift is the same as an equivariant action of the Lie
algebra).
Next, let us determine
when the extension \eqref{galgextension} splits.  Let
\begin{equation*}
   \mu\colon \: M \: \longrightarrow
 \: \mathfrak{g}^*
\end{equation*}
be the moment map.  In the notation of \eqref{MomentMaps}, the functions
\begin{equation*}
   D^{(a)} \: = \: 2 \langle\mu,t^a\rangle
\end{equation*}
for $t^a\in \mathfrak{g}$.
The condition for \eqref{galgextension} to split is that $\mu$ be equivariant,
{\it i.e.} that
\begin{equation}\label{muequivariant}
 \langle\mu,[t^a, t^b]\rangle \: = \: 
\{\langle\mu,t^a\rangle,\langle\mu,t^b\rangle\}
\end{equation}
where $\{\cdot,\cdot\}$ is the Poisson-bracket, defined using $\omega'$.
Given two splittings of \eqref{galgextension}, the \emph{difference} is an
element of ${\rm Hom}(\mathfrak{g},H^0(M,\mathbb{R}))$.

Having found an infinitesimal lift, an equivariant lift of the Lie algebra,
it remains to find a lift of the group $G$ itself, 
that is, a splitting of the exact sequence of groups
\begin{equation*}
1 \to U(1)\to \tilde{G}\to G \to 1
\end{equation*}
For $G$ connected and simply-connected, there is no further obstruction. Given a lift at the level of the Lie algebra, every path in $G$ lifts (uniquely!) to a path in $\tilde{G}$. Moreover, two paths in $G$, which are homotopic, lift to homotopic paths in $\tilde{G}$. If $G$ is not simply-connected, there's no guarantee that a \emph{closed} path in $G$ lifts to a closed path in $\tilde{G}$. But for $G$ simply-connected, every closed path is homotopic to the trivial path, and hence lifts to a closed path in $\tilde{G}$.

For $G$ not simply-connected, we may need to go to a finite
cover $\hat{G}\to G$, in order to find an equivariant lift.
As an example, consider $M= \mathbb{C}P^1$, with $G=SO(3)$.
As discussed in \cite{Bagger:1982ab}, when $\mathcal{L}$ is an odd power
of the tautological line bundle, it is $\hat{G}=SU(2)$ that lifts
to an equivariant action on $\mathcal{L}$; $\mathcal{L}$ does not admit
an $SO(3)$-equivariant structure.

As noted earlier, when equivariant structures do exist,
they are not unique.  The set of
equivariant structures on $C^{\infty}$ line bundles
preserving the connection form torsors under
${\rm Hom}(G, U(1))$; the set of equivariant structures on
holomorphic line bundles preserving the holomorphic structure
form torsors under ${\rm Hom}(G,\mathbb{C}^{\times})$.
As noted elsewhere in this paper, the Fayet-Iliopoulos parameters
correspond precisely to such choices (and hence are quantized).

\section{An example of a GIT quotient}   \label{git-ex}

In order to clarify some of the claims made in the text,
and since the technology is not widely familiar to physicists,
in this section we shall work through a very basic example of a
geometric invariant theory (GIT) quotient.
(See \cite{git,newstead,kirwan} for more information
on GIT quotients, and \cite{sotvfp}[appendix C] for additional examples.)

In principle, given a complex manifold $X$ with very ample line bundle
$L \rightarrow X$, and the action of some group $G$ on $X$
which has been lifted to a linearization on $L$, then the
GIT quotient is defined to be
\begin{displaymath}
{\rm Proj} \: \bigoplus_{n \geq 0} H^0\left( X, 
L^{\otimes n} \right)^G
\end{displaymath}
The resulting quotient is sometimes denoted $X//G$,
and depends upon the choice of linearization.
(As discussed in the text, in `typical' cases the result will
be equivalent to a symplectic quotient, for reductions on
special (`integral') coadjoint orbits.)

To clarify, let us
describe a projective space $\mathbb{P}^{n-1}$ as
$\mathbb{C}^n // \mathbb{C}^{\times}$, where the $\mathbb{C}^{\times}$ acts
with weights 1.  We write
\begin{equation}   \label{proj-ex1}
\mathbb{C}^n//\mathbb{C}^{\times} \: = \: 
{\rm Proj} \bigoplus_{p \geq 0} H^0\left(
\mathbb{C}^n, L^{\otimes p} \right)^{\mathbb{C}^{\times}}
\end{equation}
where the line bundle $L$ is necessarily $\mathcal{O}$.
The choice of linearization, the choice of equivariant structure,
can be encoded in the $\mathbb{C}^{\times}$ action on a generator,
call it $\alpha$ of
the module corresponding to $L$.
Since $L \cong \mathcal{O}$,
\begin{displaymath}
H^0\left( \mathbb{C}^n, L^{\otimes p} \right) \: = \: 
\mathbb{C}[x_1, \cdots, x_n]
\end{displaymath}
for all $p$, but the $\mathbb{C}^{\times}$ action varies.
For example, if $s \in H^0\left( \mathbb{C}^n, L^{\otimes p} \right)$
is homogeneous of degree $d$, and the generator $\alpha$ is
of weight $r$ under $\mathbb{C}^{\times}$,
then under the $\mathbb{C}^{\times}$ action,
\begin{displaymath}
s \: \mapsto \: \lambda^{d+pr} s
\end{displaymath}
where $\lambda \in \mathbb{C}^{\times}$ -- the $\lambda^d$ because $s$ is
a degree $d$ polynomial, the $\lambda^{pr}$ because of the action on the
generator.

If $r=0$,
then for all $p$ the only $\mathbb{C}^{\times}$-invariant 
sections are constants,
so we have
\begin{displaymath}
{\rm Proj} \: \bigoplus_{p \geq 0} \mathbb{C} \: = \:
{\rm Proj} \: \mathbb{C}[y]
\end{displaymath}
where $y$ is taken to have degree 1.  This is just a point.

If $r = -1$, then the $\mathbb{C}^{\times}$-invariant sections of
$L^{\otimes p}$ are homogeneous polynomials of degree $p$.
In this case, (\ref{proj-ex1}) becomes
\begin{displaymath}
{\rm Proj} \: \bigoplus_{p \geq 0} \left( \mbox{degree $p$ polynomials} \right)
\: = \: {\rm Proj} \: \mathbb{C}[x_1, \cdots, x_n]
\end{displaymath}
where the $x_i$ all have degree 1, which is $\mathbb{P}^{n-1}$.

If $r < -1$, then (\ref{proj-ex1}) becomes
\begin{displaymath}
{\rm Proj} \: \bigoplus_{p \geq 0} \left( \mbox{ degree $-rp$ polynomials}
\right)
\end{displaymath}
The map from $\mathbb{C}[x_1, \cdots, x_n]$ into the graded ring above
defines the Veronese embedding of degree $-r$ of $\mathbb{P}^{n-1}$ into
$\mathbb{P}^{(n-1 + r) \: {\rm choose} (n-1) -1}$.
This is a degree $-r$ map.

If $r > 0$, then there are no invariant sections except for constant
sections in the special case that $p=0$.  In this case, (\ref{proj-ex1})
is the empty set.

Physically, the integer $r$ corresponds to the Fayet-Iliopoulos parameter
in the supergravity theory.

What distinguishes the various values of $r$ is the K\"ahler class of the
resulting space.  In the GIT quotient construction, the K\"ahler class
is the first Chern class of some line bundle, obtained by pulling back
$\mathcal{O}(1)$ along the canonical embedding into a projective space
defined by the Proj construction.
For the linearization defined by $-r$, we have seen that the GIT quotient
(defined by the Proj of the invariant subring) is given by the projective
space $\mathbb{P}^{n-1}$
together with a natural embedding of degree $-r$ into a higher-dimensional
projective space.  Pulling back $\mathcal{O}(1)$ along such a map
gives $\mathcal{O}(-r)$.  Thus, the linearization defined by $r$ ($r < 0$)
corresponds to a K\"ahler class $-r$ on $\mathbb{P}^{n-1}$.

In principle, one would like a gerby analogue of the construction
above, that produces a closed substack of a gerbe on a projective space,
rather than a closed subvariety of a projective space.
However, we do not know of a precise analogue -- meaning,
the Proj construction always builds spaces, not stacks, and we do
not know of a stacky analogue of Proj.

\addcontentsline{toc}{section}{References}
\bibliographystyle{utphys}
\bibliography{git}

\end{document}